# Relay Divide & Rule Technique to Solve Energy Hole Problem for Wireless Sensor Network

JagpreetSingh[1], Satish Arora[2]

Computer and Science Engineering, Punjab Technical University
Jalandhar, Punjab, India

[1]p_randhawa@rocketmail.com

[2]arora3683@gmail.com

## Abstract

*In remote sensor system, sensor hubs have the restricted battery power, so to use the vitality in a more productive manner a few creator's created a few strategies, yet at the same time there is have to decrease the vitality utilization of hubs. In this paper, we presented another method called as 'Partition and Rule strategy' to unravel the scope dividing so as to open the system field into subfield and next maintain a strategic distance from the vitality gap issue with the assistance of static bunching. Essentially, in gap and run plan system range is isolated into three locales to be specific internal, center and external to conquer the issue of vitality utilization. We execute this work in NS-2 and our recreation results demonstrate that our system is far superior than old procedures.*

## 1. INTRODUCTION

Wireless Sensor Networks (WSNs) have been broadly considered as a standout amongst the most critical advances for the twenty-first century. A remote sensor system is a kind of remote system. A wireless sensor network is a wireless network that made up of a number of sensors node and at least with one base station. There are different role of Sensors in different applications. These can sense diverse natural (temperature, stickiness, light, and so on.), submerged (water saltiness, seismic checking, oil contamination observing, and so on.), and human body parameters (human crucial signs), and so forth, [2]. In wireless network a collection of nodes organized in a cooperative network. Vitality sparing is the urgent issue in planning the remote sensor systems. So as to expand the lifetime of sensor hubs, it is desirable over circulate the vitality disseminated all through the remote sensor system. Remote sensor system has two sorts organized and unstructured. In organized remote sensor arrange, the all sensor hubs are conveyed in pre planned way. In unstructured a collection of sensor nodes deployed in ad-hoc manner into a region. Once conveyed, the system is missing unattended perform checking and reporting capacities. The advantage of structure remote sensor system is that a few hubs can be conveyed with lower system upkeep and administration cost. Less hubs can be conveyed at particular areas to give scope while specially appointed arrangement can have revealed districts. Sensor hubs are gathering information about environment, in the wake of gathering it they handle it and after that transmit to the base station. Many author's try to reduce the energy consumption of sensor nodes, but the demand is still there [1]. Essential normal for the remote sensor system are restricted vitality, element system topology, lower force, hub disappointment, versatility of the hubs, short-range show correspondence, multi-jump steering and extensive size of arrangement. Energy consumption and network life time has been considered as the major issues in wireless sensor network (WSN). Firstly, placing the nodes in network is an important step, also beneficial for avoiding energy hole and coverage hole problems. Because if the first node degrade their battery then it put impact on whole area [8]. What is next, clustering is the best method to enhance the network lifetime and creating the optimum number of CH also play an crucial role, which enhance the network stability and lifetime [6].Wireless sensor networks follow some approaches for improving the lifetime of network like energy-aware technique, multi-hop routing and density control technique but these approaches still need to be improved. On the basis of network structure, routing protocol is divided into three parts: flat routing protocol, Hierarchical routing protocol and location aware [7] . In flat all the nodes have same rules i.e. nodes can sense the environment and sending the data to the base station. So it has very low network lifetime In various leveled the low vitality hubs sense the earth and high vitality hubs used to send the information to the base station. Area based steering can be utilized as a part of systems where sensor hubs have the capacity to decide their position utilizing an assortment of limitation framework and calculation.

In this exploration work, we present another bunching procedure at directing layer named as Enhanced Divide-and-Conquer Transfer Relay system (EDCTR). In this examination we concentrate on two primary targets first is scope opening and another is vitality gap [4].

## RELATED WORK

H. Dhawan et al., describe the clustering based routing protocol Low Energy Adaptive Clustering Hierarchy (LEACH) [12]; the most effective and useful protocol in WSN. It achieves efficient energy consumption results by distributing the energy equally among nodes. But because of this random selection of nodes, sometime the node which has lower power select as a CH, which further is a negative point of this technique because it may result in network failure.

W.R. Hienzelman, et al., proposed LEACH-Centralized (LEACH-C) is an augmentation of LEACH, which is propose by him [13]. The Leach-C basically uses the combined concept of routing protocol and media access to enhance the lifetime of wireless micro-sensor network. The main point here is that the BS selects the CH according to the battery power of node, which remove the drawback of LEACH.

Peyman Neamatollahi et al. [1] clarified that bunching is a compelling methodology for sorting out the system into an associated chain of command, burden adjusting, and upgrading the system lifetime. Basically, two types of clustering is used first one is static and second is dynamic. In static, once the CH is formed it can't change, whereas in Dynamic after one network operation CH is changed. This paper displays a Hybrid Clustering Approach (HCA).

Jakob Salzmann et al. presented [3] an augmentation of MASCLE convention as HEX_MASCLE, which upgrade the system lifetime by changing the state of cells. It utilize the joined idea of two stage cycle (2-MASCLE) and four stage cycle(4-MASCLE) to diminish the vitality utilization. Simulation result shows the enhanced result as compared to predecessors.

K. Latif et. al. [4] proposed a new clustering technique namely 'Divide and Conquer' to prolong the network lifetime. In this paper author firstly divide the area into three concentric squares such as inner, middle and outer and by using multi-hop communication author reduce the distance for sending information to the BS. Here author also use the mid-term algorithm to elect the CH in each region, which further collect the data from their respective sensor nodes and aggregate these all data. Next of it the aggregated data is send to the BS via multi-hop routing. Simulation results show that the energy consumption as compared to old fashioned techniques such as LEACH, LEACH-C, is far better and also show that the number of alive nodes is more.

Basilis Mamalis et.al [4] depicts the idea of Clustering and portrayed different outline difficulties of bunching in Wireless Sensor systems. The paper likewise depicts different bunching Protocols including Probabilistic Clustering Approaches and Non-Probabilistic Clustering Approaches. The calculations talked about in these conventions consider intermittently re-race of Cluster Heads (revolution of Cluster Head part) among all hubs. The fundamental downside of these calculations is that the time multifaceted nature of these calculations is hard to be kept low as the extent of the Wireless sensor Networks gets to be bigger and bigger, the augmentation in multi-bounce correspondence examples is unavoidable which expands the directing way.

Kiran Maraiya et.al [5] has exhibited a review of remote sensor system, how remote sensor systems works and different uses of remote sensor systems. In this paper it has been depicted that qualities of remote sensor system are dynamic system topology, lower force, hub disappointment and portability of hubs, short-range show correspondence and multi-jump steering and huge size of arrangement. Be that as it may, low force of sensor hubs is one of the confinements of remote sensor system as in brutal situations it is hard to supplant sensor hubs so low power may bring about vitality opening in remote sensor systems. Likewise multi-bounce steering may bring about more hubs drain their vitality while directing when contrasted with single jump directing.

## 3. PROTOCOL DESIGN

In this chapter we first focus on formation of cluster in order to solve the problem of energy hole and then divide the network into three sub-regions for coverage hole problem.

### 3.1 Cluster formation

In EDCR, static clustering is used. Nodes are randomly deployed in the each cluster and BS is located at the centre of the network. Mostly CH's elected on the basis of probability, but in our algorithm CH is selected on the basis of MID-TERM point and by using LEACH. The node which is closest to the mid-point of one region elected as CH and then by using LEACH algorithm based on energy of node, remaining CH is selected for other network operation.

Table 1. Notations used in mathematical model

| Symbol | Meaning |
|---|---|
| Is | Internal Square |
| Ms | Middle Square |
| Os | Outer Square |
| Sn | $n^{th}$ Segment |
| Sn | $n^{th}$ Sqaure |
| $T_r$ | Top right corner of internal square |
| $T_l$ | Top left corner of internal square |
| $B_r$ | Bottom right corner of internal square |
| $B_l$ | Bottom left corner of internal square |
| D | Distance |
| a | Sensor node |
| b | Relay node |
| m | Energy of relay node |
| $m_0$ | Energy of sensor node |
| u | For adding node in future |
| Ei(r) | Energy consumption of relay nodes |
| E(r) | Energy consumption of sensor nodes |

Initially, network is divided into 3 concentric squares. These squares are known as internal, middle and outer regions. Following equations divided network into concentric squares –

$$T_r (Is) (x2, y2) = (x1 + d, y1 + d), \quad (1)$$

$$B_r (Is) (x3, y3) = (x1 + d, y1 - d), \quad (2)$$

$$T_l (Is) (x4, y4) = (x1 - d, y1 + d), \text{ and} \quad (3)$$

$$B_r (Is) (x5, y5) = (x1 - d, y1 - d). \quad (4)$$

Where, d is the separation from the focal point of system field to the limit of Is. It is likewise the reference separation for isolating the entire system territory into concentric squares.

On the off chance that we have n number of concentric squares then the corner directions of Sn can be ascertained as:

$T_r (S_n) (x_n, y_n) = (x_1 + d_n, y_1 + d_n),$ (5)

$B_l (S_n) (x_n, y_n) = (x_1 + d_n, y_1 - d_n),$ (6)

$T_l (S_n) (x_n, y_n) = (x_1 - d_n, y_1 + d_n),$ and (7)

$B_r (S_n) (x_n, y_n) = (x_1 - d_n, y_1 - d_n).$ (8)

After deploying the sensor nodes we put relay nodes inside the internal region to prolong the lifetime of network. Since relay hub has high power as contrast with sensor hub and it's likewise skilled for exchanging the immense measure of information. The system that we use to include the transfer hubs into the system is –

Popt (1+b) Ei(r) / (1+m (a+m0 (-a+b +m (-b+u))))) E(r)

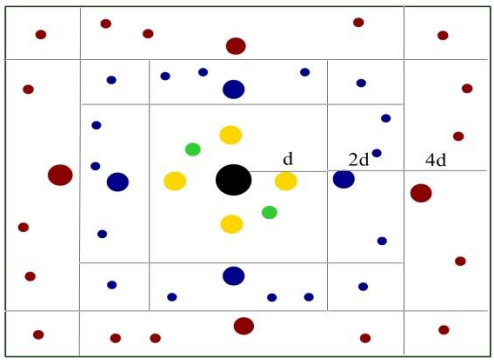

Figure 1. Deployed area

The inner region like where the yellow node is replaced having no CH and these nodes consumes more energy because it handle the data of outer as well as middle region. In order to reduce the consumption of nodes we put the relay node(green color) inside the internal square.

## 4. SIMULATION RESULTS

With the help of Network Simulation (NS-2) we generated the network with 41 nodes. The simulation result has been taken out in the NS-2 tool. There are number of packets shown on y-axis and time is given on x-axis in seconds.

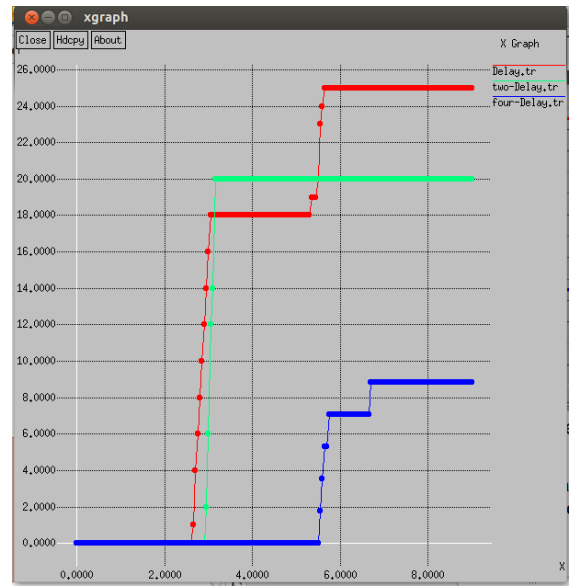

Figure 2. Delay

The figure 2 shows the experimental results of Delay of packets. Here we compare the result of Delay with the network of only sensor nodes, sensor nodes along with two relay nodes and sensor nodes with four relay nodes. Finally the network with four relay nodes has less delay than other networks.

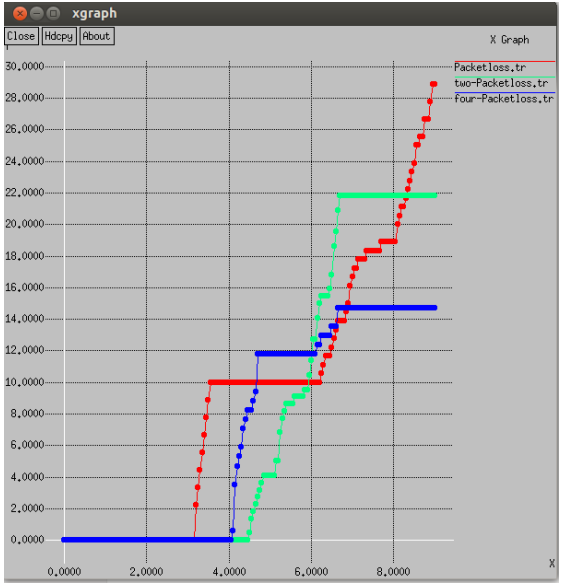

Figure 3. Packet Loss

The figure 3 shows the experimental results of Packet loss. Same as delay here we also compare the results with simple network, network with two relay nodes and four nodes and the result of four relay nodes again high.

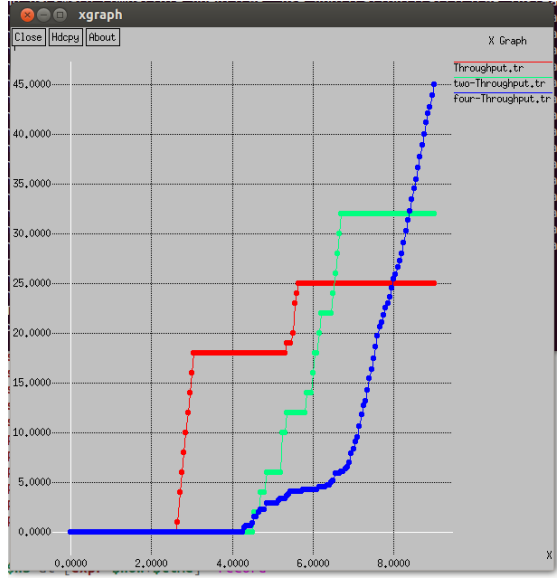

Figure 4. Throughput

The figure 4 shows the experimental results of throughput. With the availability of more relay nodes such as four relay nodes, hence, the range of packets transmission is more for that network.

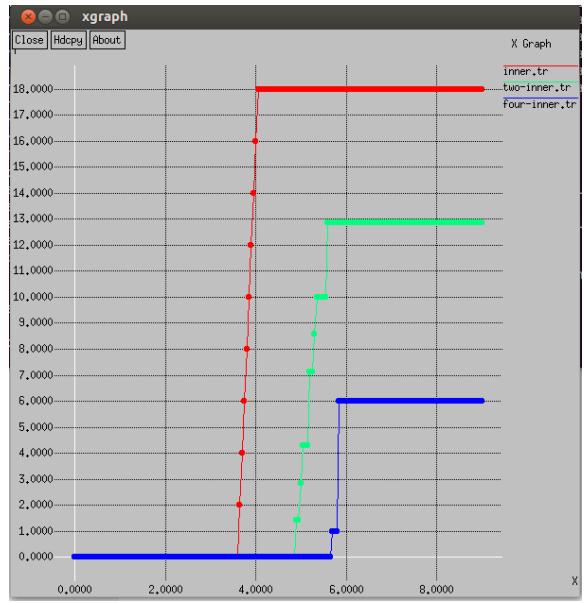

Figure 5. Energy consumption of Inner region

We show the enhanced result than old fashioned techniques. Now the internal square consumes less energy because of the less energy load on sensor nodes and network lifetime is extended.

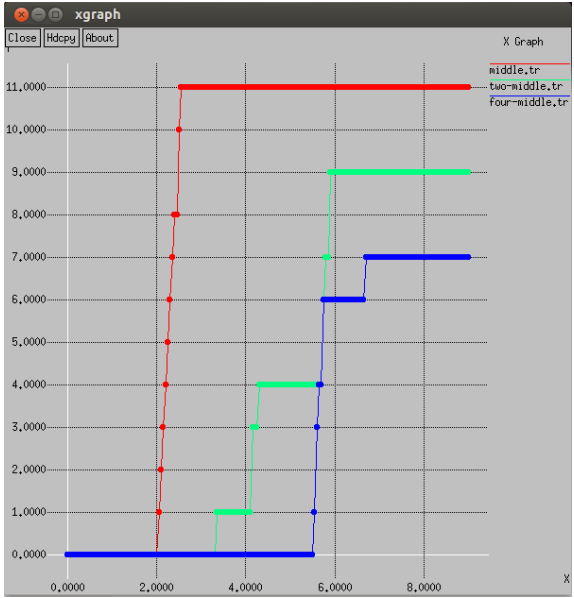

Figure 6. Energy Consumption of Middle Region

Similar to inner region energy consumption, the middle regions also show the enhanced results as compared to previous techniques such as LEACH.

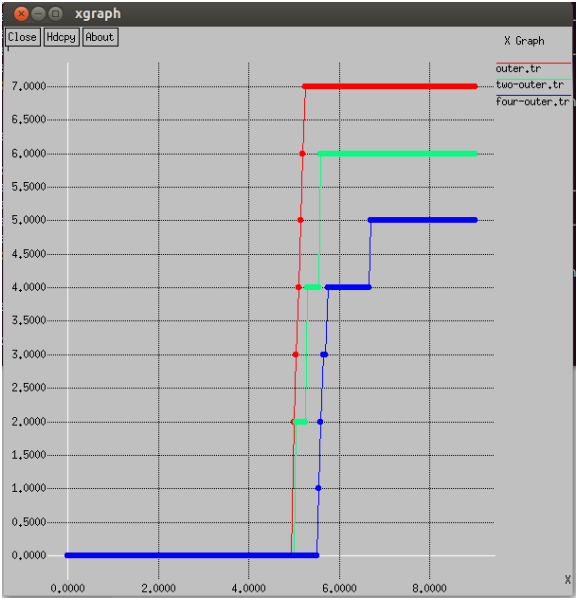

Figure7. Energy Consumption of Outer region.

With the availability of relay nodes in the network, we show that the network lifetime of all regions are extended.

## 5. CONCLUSION

In this article, our main objective is energy efficient routing, and by using static clustering as well as the concept of dividing the network into sub-regions we reduce the energy consumption. Moreover, multi-hop communications also reduce the distance between nodes and regions. By selecting the CH on the basis of mid-term point and LEACH we show that our technique produced more effective results in terms of energy, throughput, delay and packet loss. In future, we may replace the inner region sensor nodes with relay nodes to extend the lifetime of network.

## AUTHORS

Jagpreet Singh is pursuing M.TECH final year in department of Computer and Science Engineering at CT Institute of Technology and Research, Jalandhar. He has done his B.TECH in trade Computer and Science engineering from SGI college of Engineering and Technology. He has presented many papers in national and international conferences. His topic of research is Prolong the Network Stability.

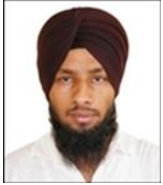